\documentclass[%
reprint,
superscriptaddress,
nofootinbib,
 amsmath,amssymb,
 aps,
 showpacs,
 physrev,
]{revtex4-1}

\usepackage[unicode=true,pdfusetitle,
 bookmarks=true,bookmarksnumbered=false,bookmarksopen=false,
 breaklinks=false,pdfborder={0 0 1},backref=false,colorlinks=true]{hyperref}
\hypersetup{
 urlcolor=blue, citecolor=blue}
\usepackage{amsmath}
\usepackage{tabularx}
\usepackage{footnote}

\usepackage[pdftex]{color}
\usepackage[pdftex]{graphicx}
\usepackage{dcolumn}
\usepackage{bm}
\usepackage{lineno}

\usepackage[caption=false]{subfig}

\usepackage{threeparttable}

\usepackage[normalem]{ulem} 
\begin{document}

\preprint{APS/123-QED}

\title{Precise Mass Measurement of the \texorpdfstring{$^{149}$La-$^{149}$Ce-$^{149}$Pr isobaric chain}{149La, 149Ce, 149Pr} }%

\author{B.~Liu}
\affiliation{Department of Physics and Astronomy, University of Notre Dame, Notre Dame, IN 46556, USA}
\affiliation{ Physics Division, Argonne National Laboratory, Lemont, IL 60439, USA}

\author{M.~Brodeur}
\email{mbrodeur@nd.edu}
\affiliation{Department of Physics and Astronomy, University of Notre Dame, Notre Dame, IN 46556, USA}

\author{J.A.~Clark}
\affiliation{ Physics Division, Argonne National Laboratory, Lemont, IL 60439, USA}

\author{D.~Ray}
\affiliation{Department of Physics and Astronomy, University of Manitoba, Winnipeg, MB R3T 2N2, Canada }
\affiliation{ Physics Division, Argonne National Laboratory, Lemont, IL 60439, USA}

\author{G.~Savard}
\affiliation{ Physics Division, Argonne National Laboratory, Lemont, IL 60439, USA}
\affiliation{Department of Physics, University of Chicago, Chicago, IL 60637, USA}

\author{A.A.~Valverde}
\affiliation{Department of Physics and Astronomy, University of Manitoba, Winnipeg, MB R3T 2N2, Canada }
\affiliation{ Physics Division, Argonne National Laboratory, Lemont, IL 60439, USA}

\author{D.P.~Burdette}
\affiliation{ Physics Division, Argonne National Laboratory, Lemont, IL 60439, USA}

\author{A.M.~Houff}
\affiliation{Department of Physics and Astronomy, University of Notre Dame, Notre Dame, IN 46556, USA}

\author{A.~Mitra}
\affiliation{Department of Physics and Astronomy, University of Notre Dame, Notre Dame, IN 46556, USA}

\author{G.E.~Morgan}
\affiliation{Department of Physics and Astronomy, Louisiana State University, Baton Rouge, LA 70803, USA}
\affiliation{ Physics Division, Argonne National Laboratory, Lemont, IL 60439, USA}

\author{R.~Orford}
\affiliation{Nuclear Science Division, Lawrence Berkeley National Laboratory, Berkeley, California 94720, USA}

\author{W.S.~Porter}
\affiliation{Department of Physics and Astronomy, University of Notre Dame, Notre Dame, IN 46556, USA}

\author{C.~Quick}
\affiliation{Department of Physics and Astronomy, University of Notre Dame, Notre Dame, IN 46556, USA}

\author{F.~Rivero}
\affiliation{Department of Physics and Astronomy, University of Notre Dame, Notre Dame, IN 46556, USA}

\author{K.S.~Sharma}
\affiliation{Department of Physics and Astronomy, University of Manitoba, Winnipeg, MB R3T 2N2, Canada }

\author{L.~Varriano}
\thanks{Present address: Center for Experimental Nuclear Physics and Astrophysics, University of Washington, Seattle, WA 98195}
\affiliation{Department of Physics, University of Chicago, Chicago, IL 60637, USA}
\affiliation{ Physics Division, Argonne National Laboratory, Lemont, IL 60439, USA}

\date{\today}
\begin{abstract}

Penning trap mass measurements of $^{149}$La, $^{149}$Ce, and $^{149}$Pr were performed with the Canadian Penning Trap (CPT) at the CARIBU facility of Argonne National Laboratory using the phase-imaging ion-cyclotron-resonance technique. The resulting mass excess of $^{149}$La differs by 221 keV from a recent JYFLTRAP measurement, resulting in a significant change in the profile of the two-neutron separation energy for that isotopic chain. The mass excesses of $^{149}$Ce and $^{149}$Pr are determined with an eight-fold improvement in precision compared to previous time-of-flight ion-cyclotron-resonance measurements; the $^{149}$Ce value is consistent with AME2020, while the $^{149}$Pr mass excess is lower by 17.5 keV. The mass excesses of $^{149}$La and $^{149}$Pr reported in this work have been confirmed recently by a measurement with a multi-reflection time-of-flight mass spectrometer coupled to a $\beta$-time of flight detector at RIKEN, providing further validation of the present results.

\end{abstract}

\maketitle

\section{Introduction}

Over the past decades, atomic masses have been used to uncover changes in nuclear structure from the vanishing of the $N=20$ nuclear shell in neutron-rich Na isotopes, reported in the seminal work of Thibault et al. \cite{Thibault75}, to the discovery of the $N = 32$ shell closure in Ca isotopes at ISOLTRAP \cite{Wienholtz2013}. These changes in nuclear structure were found by inspecting the two-neutron separation energy $S_{2n}$ of neighboring isotopic chains where a shell closure is followed by a sudden drop in $S_{2n}$-values. Likewise, that same mass metric can be used to uncover or confirm regions of rapid changes in nuclear shape such as the ones near $N=60$ \cite{Hager2006}. In such cases, the onset of deformation is indicated by a sudden increase in the $S_{2n}$-value.

More recently, high-precision mass measurements of lanthanide isotopes using the JYFLTRAP Penning trap \cite{Jaries2025} indicated a two-isotope persistence of the increased $S_{2n}$-value at $N = 93$ and 94, unlike the single jump at $N = 93$ observed for the Ce chain. In order to independently confirm this trend, high-precision mass measurements of $^{149}$La, $^{149}$Ce, and $^{149}$Pr were performed using the Canadian Penning Trap mass spectrometer at the CARIBU facility.

\section{Experiment setup}

The experiments were performed with the Canadian Penning Trap (CPT) mass spectrometer, using beams delivered from the Californium Rare Isotope Breeder Upgrade (CARIBU) facility~\cite{Savard2006,CARIBU} at Argonne National Laboratory (ANL).

The CARIBU facility employs a $^{252}$Cf fission source positioned at the entrance of a large-volume helium-filled gas catcher. This catcher collects the wide range of fission fragments and guides them toward the extraction region. Downstream, a high-resolution magnetic separator ($R=m / \Delta m\sim$10,000) removes non-isobaric contaminants~\cite{Davids08}. The remaining isobaric beam is directed into a radio frequency quadrupole (RFQ) cooler-buncher, where the ions are cooled, collected and formed into bunches. These bunches are then passed through a multi-reflection time-of-flight (MR-TOF) spectrometer, which provides mass resolving power up to 100,000 and enables removal of the isobaric contaminants~\cite{CARIBU-MRTOF}. At the MR-TOF exit, a Bradbury–Nielsen gate (BNG) is used to eliminate residual unwanted species based on their time of flight.

The purified beam is then transferred to the CPT tower. This system first consists of a Paul trap serving as a preparation trap to cool and bunch the ions prior to injection into the Penning trap, which is housed inside a large superconducting magnet at the top of the tower. Following the measurement cycle inside the Penning trap, the ions are extracted from the Penning trap and guided through a long drift tube toward a position-sensitive microchannel plate (PS-MCP) detector, positioned far enough from the magnetic field to allow proper detection. A detailed description of the CPT tower setup at CARIBU is provided in \cite{Savard2006,Ray2025}.

\section{Measurement method}

In Penning trap mass spectrometry, the mass of an ion is determined from its cyclotron frequency based on the following equation:

\begin{equation}
\nu_\mathrm{c}=\frac{qB}{2\pi m},
\label{eq:vc1}
\end{equation}

where $B$ is the magnetic field encompassing the trap and $\nu_\mathrm{c}$ is the cyclotron frequency of the ion of interest of charge $q$ and mass $m$. In a Penning trap, the ion motion consists of three eigenmotions: an axial oscillation and two radial motions -- the reduced cyclotron motion and the magnetron motion. The sum of the two radial motion frequencies yields the cyclotron frequency of the ion. The cyclotron frequency can be measured using either the time-of-flight ion-cyclotron-resonance (TOF-ICR) method~\cite{KONIG199595, Bollen:204719} or the phase-imaging ion-cyclotron-resonance (PI-ICR) \cite{PhysRevLett.110.082501} method. In recent years the CPT system has been primarily using the PI-ICR technique since it provides higher precision and allows to perfom a measurement with a lower number of ions as compared to the TOF-ICR technique. The implementation of the PI-ICR technique at the CPT is described in detail in \cite{Orford2020}.

At the CPT, in a typical PI-ICR measurement cycle, ions are first captured in the Penning trap and subsequently manipulated using a sequence of radiofrequency (RF) excitations. First, a dipole excitation at the reduced cyclotron frequency is applied to excite the ions to a larger orbital radius, after which the ions are left rotating in mainly a reduced cyclotron motion for a given amount of time $t_{\mathrm{acc}}$. Different ion species with different mass-to-charge ratios exhibit distinct reduced cyclotron frequency and accumulate different phases during this time interval. A subsequent quadrupole excitation converts the ions from the reduced cyclotron motion to a magnetron motion in preparation for ejection. After ejection, the ions are transported through a drift tube and detected by the PS-MCP detector, where their phases are recorded. In a typical measurement, two phases are recorded: a reference phase is first taken with zero accumulation time and then a final phase is taken with a chosen non-zero accumulation time. The difference between the reference phase and final phase yields the total accumulated phase difference and is used to determine the cyclotron frequency of the ion of interest via:

\begin{equation}
\nu_\mathrm{c}=\frac{\phi_{\mathrm{tot}}}{2\pi t_{\mathrm{acc}}}=\frac{\phi_{\mathrm{diff}}+2\pi N}{2\pi t_{\mathrm{acc}}},
\label{eq:vc2}
\end{equation}

where $\phi_{\mathrm{tot}}$ is total phase accumulated between the reference phase and the final phase, $\phi_{\mathrm{diff}}$ is the phase difference between the two measured phase spots, and $N$ is the integer number of turns that the final phase advanced beyond the reference phase.

In practice, rather than determining the magnetic field strength and calculating the mass directly from Eq.~\ref{eq:vc1}, the cyclotron frequency of a calibrant with a relatively well-known mass is measured. The mass of the ion of interest is then determined from the ratio of the cyclotron frequencies,

\begin{equation}
m=\frac{\nu_{\mathrm{c,cal}}}{\nu_\mathrm{c}} \frac{q}{q_{cal}}(m_{\mathrm{cal}}-q_{\mathrm{cal}}m_\mathrm{e})+qm_\mathrm{e},
\label{eq:vc2}
\end{equation}

where $m$, $q$, $\nu_\mathrm{c}$ denote the mass, charge and cyclotron frequencies. The subscript \textit{cal} refers to the calibrant. The electron mass is denoted by $m_e$. Contributions from electron binding energies can be neglected at the level of precision considered in this work.

\section{Experimental details and data analysis methods}

The mass measurements of $^{149}$La, $^{149}$Ce, and $^{149}$Pr (all doubly-charged) were performed using the CPT with CARIBU beams. Singly-charged $^{12}$C$_{6}$$^{1}$H$_{6}$ was taken as the calibrant for the measurement and it was also produced within the CARIBU system. It was chosen as a calibrant due to having a mass-to-charge ratio close to that of the ions of interest, thereby reducing mass-dependent systematic effects.

\begin{figure}[tpb]
  \begin{center}
    \centerline{\includegraphics[width=\linewidth,trim={6cm 0.5cm 7cm 1.5cm},clip]{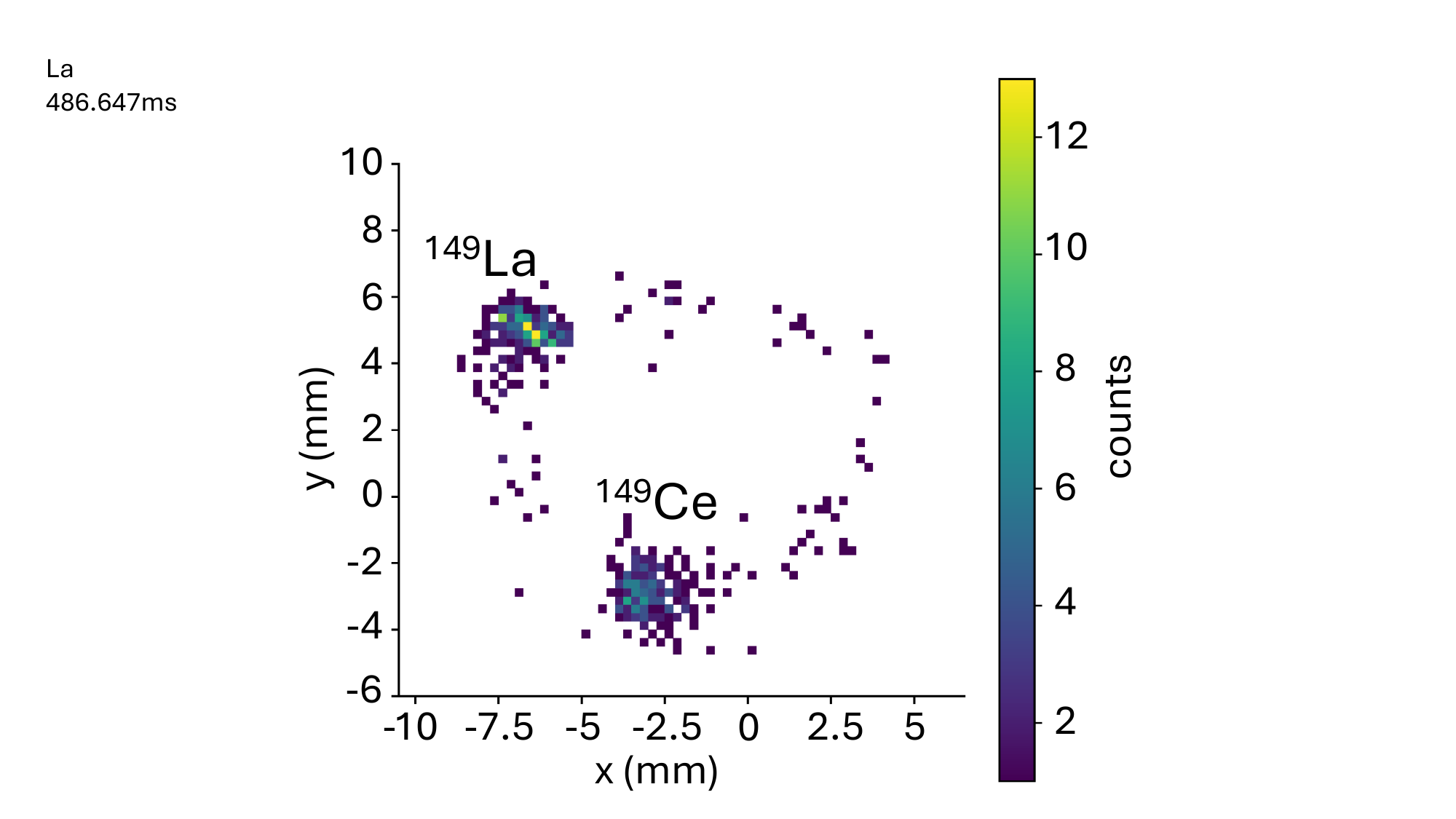}}
    \caption{Typical count histograms recorded during the measurement of $^{149}$La, where both $^{149}$La and $^{149}$Ce were present in the trap. The sparse counts distributed around are attributed primarily to background noise.}
    \label{fig:149_spots}
  \end{center}
\end{figure}

To ensure precise and accurate measurements, significant effort was dedicated to confirm the identity of the isotope of interest before any measurement. This was achieved by scanning the BNG gate timing to selectively introduce all nearby isotopes into the trap, followed by determining their cyclotron frequencies using accumulation times ranging from a few milliseconds up to several hundred milliseconds. The resulting frequency values were compared with those expected based on the AME2020 \cite{AME2020} mass excess to distinguish each isotope. In addition, the relative abundance of the observed ion species were compared against the known fission yields of $^{252}$Cf as a secondary validation. Fig.~\ref{fig:149_spots} shows a representative measurement count histogram in which both $^{149}$La and $^{149}$Ce ion spots are clearly identified during a typical $^{149}$La measurement.

Great care was taken to minimize potential systematic effects and improve accuracy starting from the measurement stage. The total ion rate was generally kept below 1.5 Hz during the measurement. This ensures that, with a 500 ms accumulation time, fewer than one ion is typically present in the trap on average, thereby minimizing ion–ion interaction effects. In addition, to reduce the systematic effect from the accumulation time, the cyclotron frequency of the ion of interest and the calibrant were measured within the same accumulation time range of 486--487 ms. Furthermore, the accumulation times were carefully chosen such that the final phase spot lands within 10 degrees of the initial reference phase spot thereby minimizing distortion effects arising from non-circular ion trajectories or PS-MCP image distortion. Finally, to suppress the effect from the system's temporal drifts, measurements of the calibrant and of the isotope of interest were performed back-to-back and under identical trap conditions.

\begin{figure}[tpb]
  \begin{center}
    \centerline{\includegraphics[width=\linewidth,trim={0cm 0cm 0cm 0cm},clip]{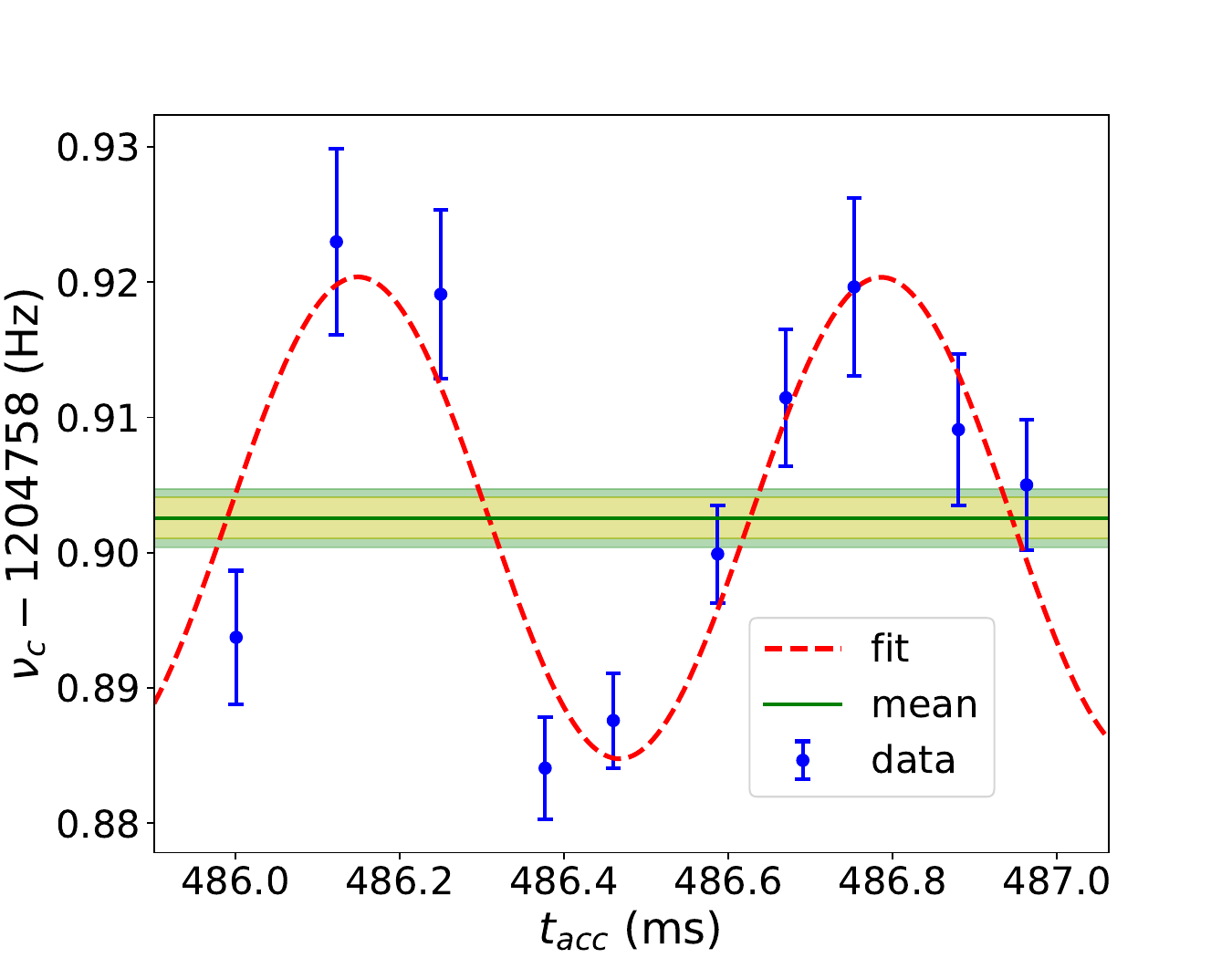}}
    \caption{Measured cyclotron frequency of $^{149}$Ce as a function of accumulation time $t_\mathrm{{acc}}$ from 486 ms to 487 ms. The dashed red curve shows a sinusoidal fit of the data. The horizontal green line marks the mean cyclotron frequency extracted from the fit. The yellow shaded band represent the fit uncertainty of the mean frequency, while the wider green band is the fit uncertainty scaled by the $\sqrt{\chi^2}$=1.41 from the fit.}
    \label{fig:149_sine}
  \end{center}
\end{figure}

Despite the precautions taken during data collection, several systematic effects remain and had to be treated during the analysis. First, the measured phase retains a dependence on the accumulation time. To characterize this behavior, the final phase was recorded multiple times at closely spaced accumulation times $t_{\mathrm{acc}}$. As described in \cite{Orford2020}, 6–7 measurements spanning at least one full magnetron period were typically acquired, resulting in a sinusoidal dependence of the extracted cyclotron frequency on $t_{\mathrm{acc}}$. As shown in Fig.~\ref{fig:149_sine}, the data were collected over $t_{\mathrm{acc}}$ from 486 to 487 ms and were fitted with a sinusoidal function, from which the mean cyclotron frequency was extracted.

In addition, a residual dependence of the cyclotron frequency on the position on the MCP detector has been observed. To account for this angular dependency, the measured frequencies were corrected using the empirical equation of angle position described in \cite{Liu2025,Liu2025Phd}. This correction resulted in a systematic downward shift of the extracted mass excesses by approximately 4–5~keV for all measured isotopes. This similar shift arises because all isotopes of interest were measured around similar positions on the detector and were calibrated using the same reference, which was at a different position.

When more than one ion species is present in the trap, the measured reference phase is a population-weighted combination of each individual reference phases rather than that of a single species. Consequently, the true reference phase of the ion of interest need be disentangled from the ensemble average. This reference-phase correction was carried out following the method of \cite{Orford2020} and with modifications described in \cite{Liu2025Phd}. For the measurement of $^{149}$La, $^{149}$Ce was the only additional species present in the trap, resulting in a correction of approximately 1~keV and an associated increase in the total uncertainty of about 30$\%$. In contrast, for the measurements of $^{149}$Ce and $^{149}$Pr, few or no contaminant species were observed, and the corresponding reference phase corrections were negligible.

Finally, the measured cyclotron frequency ratio potentially has a mass-to-charge dependent shift arising from any offset in the measured cyclotron frequency. To quantify this effect, we calculated the mass-dependent correction from a series of benchmark measurements and incorporated the magnitude of this correction as an additional systematic uncertainty in our analysis~\cite{Liu2025Phd}. While this provides an estimate of the magnitude of the mass-dependent shift, additional systematic effects have been observed that may limit the accuracy with which this correction can be reliably applied. To be conservative, the magnitude of this estimated shift in the cyclotron frequency ratio, $1.4\times10^{-9}$, was added in quadrature to the uncertainty of the frequency ratio, rather than being applied as a direct correction. Consequently, the uncertainties of the mass excesses are increased by less than 0.2 keV.

\section{Measurement results and discussion}

\begin{figure}[tpb]
  \begin{center}
    \centerline{\includegraphics[width=\linewidth,trim={0cm 0cm 0cm 0cm},clip]{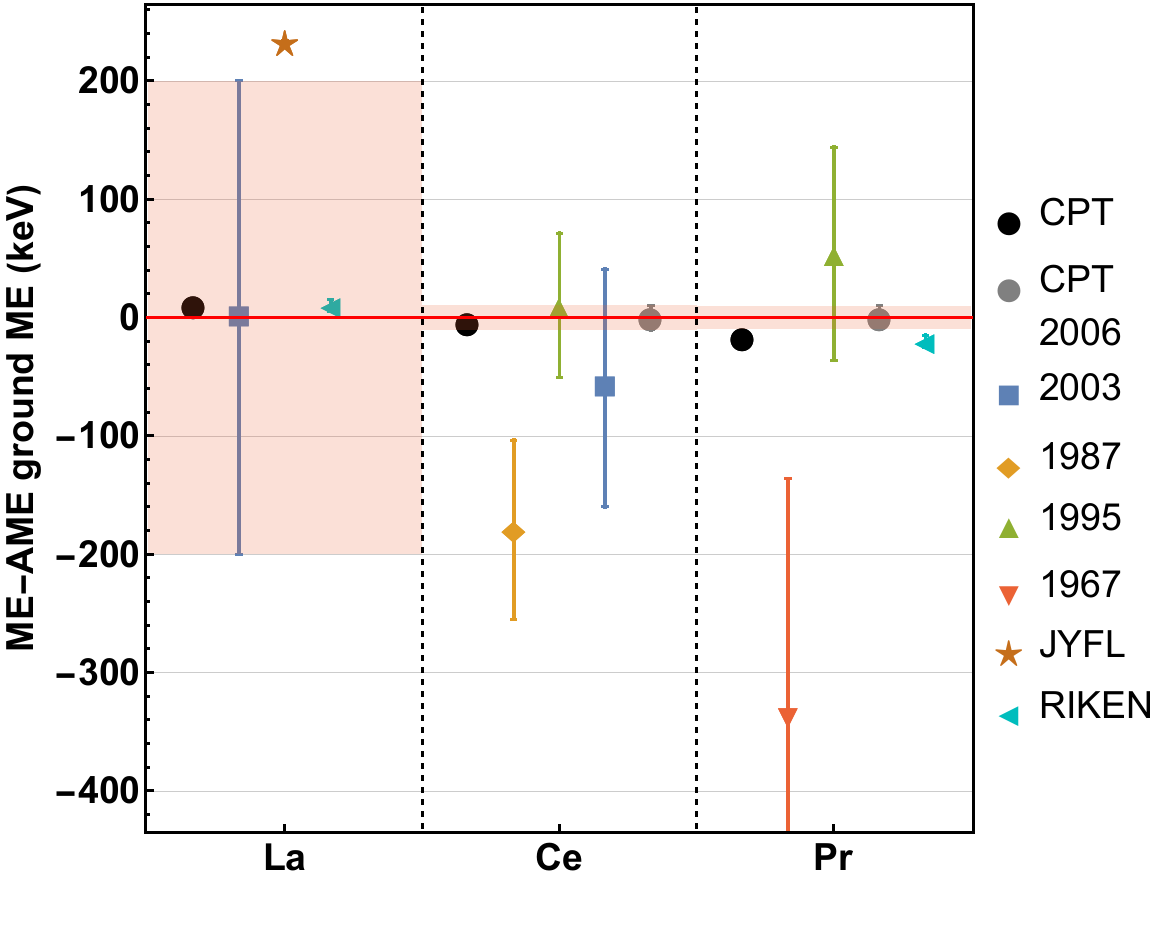}}
    \caption{Mass excess difference between this work and the AME2020~\cite{AME2020} ground state for $^{149}$La, $^{149}$Ce, and $^{149}$Pr. The solid red line and red shaded bands indicate the AME2020 ground state value and uncertainty range. Results from this work are compared with all experimental data present in AME2020:
    2006~\cite{Savard2006}, 2003~\cite{02ShB}, 1987~\cite{87GrA}, 1995~\cite{95Ik03}, 1967~\cite{67Va14}. The measurement results from JYFLTRAP~\cite{Jaries2025} and RIKEN~\cite{Kimura2025} are also shown for comparison.}
    \label{fig:149_all}
  \end{center}
\end{figure}

The cyclotron-frequency ratios following the analysis procedure described above are listed in Table~\ref{tab:CARIBU_149} together with the mass excesses calculated using the mass of the calibrant molecule, which was determined from the AME2020 value for hydrogen and ignoring molecular energies. This table also provides the AME2020 values for comparison. Also given are the mass excesses based on the recent measurements from the JYFLTRAP group \cite{Jaries2025} and using the RIKEN MR-TOF \cite{Kimura2025}. The mass excess obtained here is also shown in Fig.~\ref{fig:149_all}, alongside all previously reported experimental results considered in the AME2020 evaluation and the more recent values from JYFLTRAP and RIKEN.

As shown in Fig.~\ref{fig:149_all}, our $^{149}$La value is significantly more precise than the AME2020 while agreeing with it. It is also 221.6 keV more bound than the recent JYFLTRAP measurement. Such discrepancy is concerning, warranting further investigation. Fortunately, during the preparation of this manuscript, a mass measurement using an MR-TOF at RIKEN in combination with a $\beta$-TOF detector \cite{Kimura2025}, unambiguously confirmed our result.

The mass excess of $^{149}$Ce from our work is also consistent with the AME2020 value, based on a measurement by the CPT \cite{Savard2006} using the TOF-ICR technique, while being eight times more precise. Fig.~\ref{fig:149_all} also shows previous $\beta$-end point measurements in the AME2020 from which the mass excess of $^{149}$Ce could be derived. Good agreement is reached with these measurements with the exception of the 1987 result.

Likewise, for $^{149}$Pr, our mass excess is about eight times more precise than the AME2020 value derived from a CPT measurement using the TOF-ICR technique. However, the mass excess obtained in this work is lower than the AME2020 value by 17.5 keV. This small discrepancy of about 1.8$\sigma$ has also been resolved by a MR-TOF measurement by the RIKEN group \cite{Kimura2025}, which agrees with the new measurement result from this work.

\begin{table*}
\caption{\label{tab:CARIBU_149}%
Cyclotron frequency ratio and mass excess (ME) of the $^{149}$La, $^{149}$Ce, $^{149}$Pr measured by the CPT. The mass excess from the AME2020 \cite{AME2020}, ME$_{\textrm{AME}}$, from JYFLTRAP \cite{Jaries2025}, ME$_{\textrm{JYFL}}$, and the RIKEN MR-TOF \cite{Kimura2025}, ME$_{\textrm{RIKEN}}$ are also listed. All the isotopes of interest were in the 2+ charge state. The calibrant $^{12}$C$_{6}$$^{1}$H$_{6}$ was in 1+ charge state.}
\begin{ruledtabular}
\begin{tabular}{ccccccc}
\textrm{Nuclide}&
\textrm{Calibration}&
\textrm{$\nu_c^{cal} / \nu_c$}&
\textrm{ME(keV)}&
\textrm{ME$_{\textrm{AME}}$(keV)}&
\textrm{ME$_{\textrm{JYFL}}$(keV)}&
\textrm{ME$_{\textrm{RIKEN}}$(keV)}\\
\colrule

$^{149}$La & $^{12}$C$_{6}$$^{1}$H$_{6}$ &0.954139215(11)& -60209.9(16) & -60220(200) & -59988.3(28) & -60209.7(50) \\
$^{149}$Ce & $^{12}$C$_{6}$$^{1}$H$_{6}$ &0.9540947582(84)& -66674.0(12) & -66670(10) & &  \\
$^{149}$Pr & $^{12}$C$_{6}$$^{1}$H$_{6}$ &0.9540646172(89)& -71056.5(13) & -71039(10) & & -71059.5(52) \\

\end{tabular}
\end{ruledtabular}
\end{table*}

\begin{figure}[tpb]
  \begin{center}
    \centerline{\includegraphics[width=\linewidth,trim={0cm 0cm 0cm 0cm},clip]{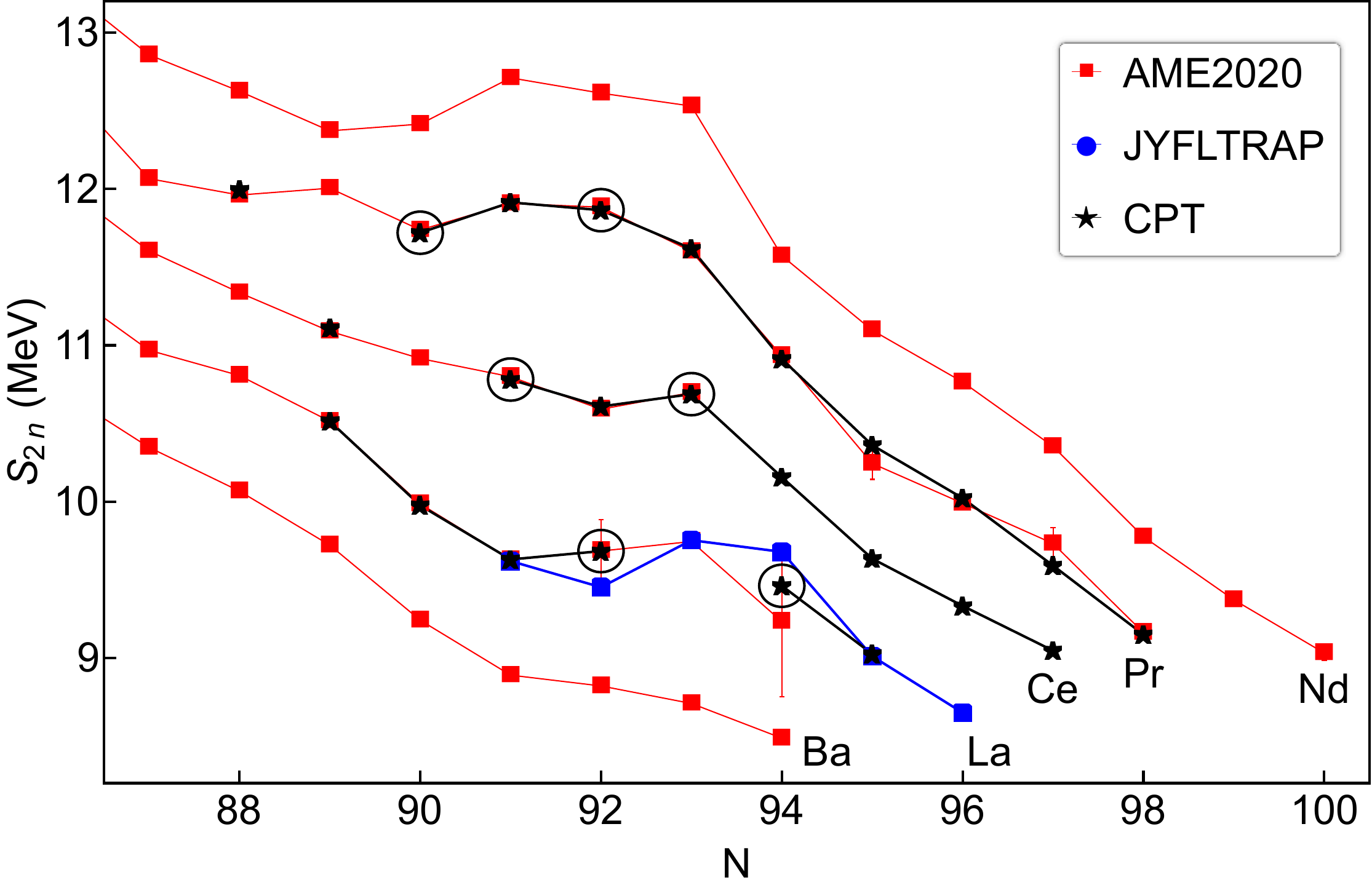}}
    \caption{Two-neutron separation energies $S_{2n}$ for the isotopic chains of Ba, La, Ce, Pr, and Nd from the AME2020 \cite{AME2020}. Also shown are recent JYFLTRAP results for the La chain, and CPT results \cite{Orford2019Phd,Orford2022,Ray2024} for the La, Ce and Pr chains. $S_{2n}$-values affected by measurements from this work are circled.}
    \label{fig:S2N}
  \end{center}
\end{figure}

The two-neutron separation energies of isotopes affected by the mass measurement of $^{149}$La, $^{149}$Ce, and $^{149}$Pr from this work are encircled in   Fig.~\ref{fig:S2N}. Shown in black are new $S_{2n}$ values for the La, Ce, and Pr isotopic chains based on all mass measurements performed with the CPT at CARIBU \cite{Orford2019Phd,Orford2022,Ray2024} since the release of the AME2020. Measurements with the CPT added four new $S_{2n}$ values to the Ce chain. In the Pr chain, the $S_{2n}$ values of $^{154}$Pr and $^{156}$Pr are the most affected, both due to the mass measurement of $^{154}$Pr that was 140.7 keV more bound that the AME2020 value based on a single $\beta$ endpoint measurement. The other significant change in $S_{2n}$ values due to a mass measurement from the CPT is observed for $^{149}$La and $^{151}$La against the recent JYFLTRAP results. Both of these changes are due to the $^{149}$La mass value from this work. Using the CPT mass excesses, the previously seen two-isotope-long jump in $S_{2n}$ is now only concentrated to $^{150}$La resulting in a $S_{2n}$ profile closer to the one seen in the Ce chain.

\section{Conclusion}

In this work, we report high precision mass measurements of $^{149}$La, $^{149}$Ce, and $^{149}$Pr, all measured using the PI-ICR technique with the CPT at the CARIBU facility. The mass excess of $^{149}$La is 221 keV lower than the previously measured value by JYFLTRAP \cite{Jaries2025}. Our result has been confirmed in a recent study, published during the writing of this manuscript, using the RIKEN MR-TOF \cite{Kimura2025} where they measured both its mass and half-life. We also report the eight times more precise masses for $^{149}$Ce and $^{149}$Pr then the AME2020 values based on measurements made using the TOF-ICR technique with the CPT. While the $^{149}$Ce mass excess agrees with the AME2020 value, the mass excess of $^{149}$Pr is lower by 17.5~keV (or 1.8$\sigma$). The value of $^{149}$Pr from this work been confirmed by a recent MR-TOF measurement by the same RIKEN group as $^{149}$La \cite{Kimura2025}. Together, these results establish an improved and independently validated mass surface for the A=149 isobaric chain.

\section*{\label{sec:ac}Acknowledgments}

This work is supported in part by the National Science Foundation under Grant No. PHY-2310059; by the U.S. Department of Energy, Office of Nuclear Physics, under Contract No. DE-AC02-06CH11357; by NSERC (Canada), under Grant No. SAPPJ-2018-00028; by the University of Notre Dame; and with resources of ANL, the ATLAS facility, and Office of Science User Facility.


\begin{thebibliography}{24}%
\makeatletter
\providecommand \@ifxundefined [1]{%
 \@ifx{#1\undefined}
}%
\providecommand \@ifnum [1]{%
 \ifnum #1\expandafter \@firstoftwo
 \else \expandafter \@secondoftwo
 \fi
}%
\providecommand \@ifx [1]{%
 \ifx #1\expandafter \@firstoftwo
 \else \expandafter \@secondoftwo
 \fi
}%
\providecommand \natexlab [1]{#1}%
\providecommand \enquote  [1]{``#1''}%
\providecommand \bibnamefont  [1]{#1}%
\providecommand \bibfnamefont [1]{#1}%
\providecommand \citenamefont [1]{#1}%
\providecommand \href@noop [0]{\@secondoftwo}%
\providecommand \href [0]{\begingroup \@sanitize@url \@href}%
\providecommand \@href[1]{\@@startlink{#1}\@@href}%
\providecommand \@@href[1]{\endgroup#1\@@endlink}%
\providecommand \@sanitize@url [0]{\catcode `\\12\catcode `\$12\catcode
  `\&12\catcode `\#12\catcode `\^12\catcode `\_12\catcode `\%12\relax}%
\providecommand \@@startlink[1]{}%
\providecommand \@@endlink[0]{}%
\providecommand \url  [0]{\begingroup\@sanitize@url \@url }%
\providecommand \@url [1]{\endgroup\@href {#1}{\urlprefix }}%
\providecommand \urlprefix  [0]{URL }%
\providecommand \Eprint [0]{\href }%
\providecommand \doibase [0]{http://dx.doi.org/}%
\providecommand \selectlanguage [0]{\@gobble}%
\providecommand \bibinfo  [0]{\@secondoftwo}%
\providecommand \bibfield  [0]{\@secondoftwo}%
\providecommand \translation [1]{[#1]}%
\providecommand \BibitemOpen [0]{}%
\providecommand \bibitemStop [0]{}%
\providecommand \bibitemNoStop [0]{.\EOS\space}%
\providecommand \EOS [0]{\spacefactor3000\relax}%
\providecommand \BibitemShut  [1]{\csname bibitem#1\endcsname}%
\let\auto@bib@innerbib\@empty
\bibitem [{\citenamefont {Thibault}\ \emph {et~al.}(1975)\citenamefont
  {Thibault}, \citenamefont {Klapisch}, \citenamefont {Rigaud}, \citenamefont
  {Poskanzer}, \citenamefont {Prieels}, \citenamefont {Lessard},\ and\
  \citenamefont {Reisdorf}}]{Thibault75}%
  \BibitemOpen
  \bibfield  {author} {\bibinfo {author} {\bibfnamefont {C.}~\bibnamefont
  {Thibault}}, \bibinfo {author} {\bibfnamefont {R.}~\bibnamefont {Klapisch}},
  \bibinfo {author} {\bibfnamefont {C.}~\bibnamefont {Rigaud}}, \bibinfo
  {author} {\bibfnamefont {A.~M.}\ \bibnamefont {Poskanzer}}, \bibinfo {author}
  {\bibfnamefont {R.}~\bibnamefont {Prieels}}, \bibinfo {author} {\bibfnamefont
  {L.}~\bibnamefont {Lessard}}, \ and\ \bibinfo {author} {\bibfnamefont
  {W.}~\bibnamefont {Reisdorf}},\ }\href {\doibase 10.1103/PhysRevC.12.644}
  {\bibfield  {journal} {\bibinfo  {journal} {Phys. Rev. C}\ }\textbf {\bibinfo
  {volume} {12}},\ \bibinfo {pages} {644} (\bibinfo {year} {1975})}\BibitemShut
  {NoStop}%
\bibitem [{\citenamefont {Wienholtz}\ \emph {et~al.}(2013)\citenamefont
  {Wienholtz}, \citenamefont {Beck}, \citenamefont {Blaum}, \citenamefont
  {Borgmann}, \citenamefont {Breitenfeldt}, \citenamefont {Cakirli},
  \citenamefont {George}, \citenamefont {Herfurth}, \citenamefont {Holt},
  \citenamefont {Kowalska}, \citenamefont {Kreim}, \citenamefont {Lunney},
  \citenamefont {Manea}, \citenamefont {Men{\'e}ndez}, \citenamefont
  {Neidherr}, \citenamefont {Rosenbusch}, \citenamefont {Schweikhard},
  \citenamefont {Schwenk}, \citenamefont {Simonis}, \citenamefont {Stanja},
  \citenamefont {Wolf},\ and\ \citenamefont {Zuber}}]{Wienholtz2013}%
  \BibitemOpen
  \bibfield  {author} {\bibinfo {author} {\bibfnamefont {F.}~\bibnamefont
  {Wienholtz}}, \bibinfo {author} {\bibfnamefont {D.}~\bibnamefont {Beck}},
  \bibinfo {author} {\bibfnamefont {K.}~\bibnamefont {Blaum}}, \bibinfo
  {author} {\bibfnamefont {C.}~\bibnamefont {Borgmann}}, \bibinfo {author}
  {\bibfnamefont {M.}~\bibnamefont {Breitenfeldt}}, \bibinfo {author}
  {\bibfnamefont {R.~B.}\ \bibnamefont {Cakirli}}, \bibinfo {author}
  {\bibfnamefont {S.}~\bibnamefont {George}}, \bibinfo {author} {\bibfnamefont
  {F.}~\bibnamefont {Herfurth}}, \bibinfo {author} {\bibfnamefont {J.~D.}\
  \bibnamefont {Holt}}, \bibinfo {author} {\bibfnamefont {M.}~\bibnamefont
  {Kowalska}}, \bibinfo {author} {\bibfnamefont {S.}~\bibnamefont {Kreim}},
  \bibinfo {author} {\bibfnamefont {D.}~\bibnamefont {Lunney}}, \bibinfo
  {author} {\bibfnamefont {V.}~\bibnamefont {Manea}}, \bibinfo {author}
  {\bibfnamefont {J.}~\bibnamefont {Men{\'e}ndez}}, \bibinfo {author}
  {\bibfnamefont {D.}~\bibnamefont {Neidherr}}, \bibinfo {author}
  {\bibfnamefont {M.}~\bibnamefont {Rosenbusch}}, \bibinfo {author}
  {\bibfnamefont {L.}~\bibnamefont {Schweikhard}}, \bibinfo {author}
  {\bibfnamefont {A.}~\bibnamefont {Schwenk}}, \bibinfo {author} {\bibfnamefont
  {J.}~\bibnamefont {Simonis}}, \bibinfo {author} {\bibfnamefont
  {J.}~\bibnamefont {Stanja}}, \bibinfo {author} {\bibfnamefont {R.~N.}\
  \bibnamefont {Wolf}}, \ and\ \bibinfo {author} {\bibfnamefont
  {K.}~\bibnamefont {Zuber}},\ }\href {\doibase 10.1038/nature12226} {\bibfield
   {journal} {\bibinfo  {journal} {Nature}\ }\textbf {\bibinfo {volume}
  {498}},\ \bibinfo {pages} {346} (\bibinfo {year} {2013})}\BibitemShut
  {NoStop}%
\bibitem [{\citenamefont {Hager}\ \emph {et~al.}(2006)\citenamefont {Hager},
  \citenamefont {Eronen}, \citenamefont {Hakala}, \citenamefont {Jokinen},
  \citenamefont {Kolhinen}, \citenamefont {Kopecky}, \citenamefont {Moore},
  \citenamefont {Nieminen}, \citenamefont {Oinonen}, \citenamefont
  {Rinta-Antila}, \citenamefont {Szerypo},\ and\ \citenamefont
  {\"Ayst\"o}}]{Hager2006}%
  \BibitemOpen
  \bibfield  {author} {\bibinfo {author} {\bibfnamefont {U.}~\bibnamefont
  {Hager}}, \bibinfo {author} {\bibfnamefont {T.}~\bibnamefont {Eronen}},
  \bibinfo {author} {\bibfnamefont {J.}~\bibnamefont {Hakala}}, \bibinfo
  {author} {\bibfnamefont {A.}~\bibnamefont {Jokinen}}, \bibinfo {author}
  {\bibfnamefont {V.~S.}\ \bibnamefont {Kolhinen}}, \bibinfo {author}
  {\bibfnamefont {S.}~\bibnamefont {Kopecky}}, \bibinfo {author} {\bibfnamefont
  {I.}~\bibnamefont {Moore}}, \bibinfo {author} {\bibfnamefont
  {A.}~\bibnamefont {Nieminen}}, \bibinfo {author} {\bibfnamefont
  {M.}~\bibnamefont {Oinonen}}, \bibinfo {author} {\bibfnamefont
  {S.}~\bibnamefont {Rinta-Antila}}, \bibinfo {author} {\bibfnamefont
  {J.}~\bibnamefont {Szerypo}}, \ and\ \bibinfo {author} {\bibfnamefont
  {J.}~\bibnamefont {\"Ayst\"o}},\ }\href {\doibase
  10.1103/PhysRevLett.96.042504} {\bibfield  {journal} {\bibinfo  {journal}
  {Phys. Rev. Lett.}\ }\textbf {\bibinfo {volume} {96}},\ \bibinfo {pages}
  {042504} (\bibinfo {year} {2006})}\BibitemShut {NoStop}%
\bibitem [{\citenamefont {Jaries}\ \emph {et~al.}(2025)\citenamefont {Jaries},
  \citenamefont {Stryjczyk}, \citenamefont {Kankainen}, \citenamefont {Eronen},
  \citenamefont {Beliuskina}, \citenamefont {Dickel}, \citenamefont {Flayol},
  \citenamefont {Ge}, \citenamefont {Hukkanen}, \citenamefont {Mougeot},
  \citenamefont {Nikas}, \citenamefont {Pohjalainen}, \citenamefont {Raggio},
  \citenamefont {Reponen}, \citenamefont {Ruotsalainen},\ and\ \citenamefont
  {Virtanen}}]{Jaries2025}%
  \BibitemOpen
  \bibfield  {author} {\bibinfo {author} {\bibfnamefont {A.}~\bibnamefont
  {Jaries}}, \bibinfo {author} {\bibfnamefont {M.}~\bibnamefont {Stryjczyk}},
  \bibinfo {author} {\bibfnamefont {A.}~\bibnamefont {Kankainen}}, \bibinfo
  {author} {\bibfnamefont {T.}~\bibnamefont {Eronen}}, \bibinfo {author}
  {\bibfnamefont {O.}~\bibnamefont {Beliuskina}}, \bibinfo {author}
  {\bibfnamefont {T.}~\bibnamefont {Dickel}}, \bibinfo {author} {\bibfnamefont
  {M.}~\bibnamefont {Flayol}}, \bibinfo {author} {\bibfnamefont
  {Z.}~\bibnamefont {Ge}}, \bibinfo {author} {\bibfnamefont {M.}~\bibnamefont
  {Hukkanen}}, \bibinfo {author} {\bibfnamefont {M.}~\bibnamefont {Mougeot}},
  \bibinfo {author} {\bibfnamefont {S.}~\bibnamefont {Nikas}}, \bibinfo
  {author} {\bibfnamefont {I.}~\bibnamefont {Pohjalainen}}, \bibinfo {author}
  {\bibfnamefont {A.}~\bibnamefont {Raggio}}, \bibinfo {author} {\bibfnamefont
  {M.}~\bibnamefont {Reponen}}, \bibinfo {author} {\bibfnamefont
  {J.}~\bibnamefont {Ruotsalainen}}, \ and\ \bibinfo {author} {\bibfnamefont
  {V.}~\bibnamefont {Virtanen}},\ }\href {\doibase
  10.1103/PhysRevLett.134.042501} {\bibfield  {journal} {\bibinfo  {journal}
  {Phys. Rev. Lett.}\ }\textbf {\bibinfo {volume} {134}},\ \bibinfo {pages}
  {042501} (\bibinfo {year} {2025})}\BibitemShut {NoStop}%
\bibitem [{\citenamefont {Savard}\ \emph {et~al.}(2006)\citenamefont {Savard},
  \citenamefont {Wang}, \citenamefont {Sharma}, \citenamefont {Sharma},
  \citenamefont {Clark}, \citenamefont {Boudreau}, \citenamefont {Buchinger},
  \citenamefont {Crawford}, \citenamefont {Greene}, \citenamefont {Gulick},
  \citenamefont {Hecht}, \citenamefont {Lee}, \citenamefont {Levand},
  \citenamefont {Scielzo}, \citenamefont {Trimble}, \citenamefont {Vaz},\ and\
  \citenamefont {Zabransky}}]{Savard2006}%
  \BibitemOpen
  \bibfield  {author} {\bibinfo {author} {\bibfnamefont {G.}~\bibnamefont
  {Savard}}, \bibinfo {author} {\bibfnamefont {J.}~\bibnamefont {Wang}},
  \bibinfo {author} {\bibfnamefont {K.}~\bibnamefont {Sharma}}, \bibinfo
  {author} {\bibfnamefont {H.}~\bibnamefont {Sharma}}, \bibinfo {author}
  {\bibfnamefont {J.}~\bibnamefont {Clark}}, \bibinfo {author} {\bibfnamefont
  {C.}~\bibnamefont {Boudreau}}, \bibinfo {author} {\bibfnamefont
  {F.}~\bibnamefont {Buchinger}}, \bibinfo {author} {\bibfnamefont
  {J.}~\bibnamefont {Crawford}}, \bibinfo {author} {\bibfnamefont
  {J.}~\bibnamefont {Greene}}, \bibinfo {author} {\bibfnamefont
  {S.}~\bibnamefont {Gulick}}, \bibinfo {author} {\bibfnamefont
  {A.}~\bibnamefont {Hecht}}, \bibinfo {author} {\bibfnamefont
  {J.}~\bibnamefont {Lee}}, \bibinfo {author} {\bibfnamefont {A.}~\bibnamefont
  {Levand}}, \bibinfo {author} {\bibfnamefont {N.}~\bibnamefont {Scielzo}},
  \bibinfo {author} {\bibfnamefont {W.}~\bibnamefont {Trimble}}, \bibinfo
  {author} {\bibfnamefont {J.}~\bibnamefont {Vaz}}, \ and\ \bibinfo {author}
  {\bibfnamefont {B.}~\bibnamefont {Zabransky}},\ }\href {\doibase
  https://doi.org/10.1016/j.ijms.2006.01.047} {\bibfield  {journal} {\bibinfo
  {journal} {International Journal of Mass Spectrometry}\ }\textbf {\bibinfo
  {volume} {251}},\ \bibinfo {pages} {252} (\bibinfo {year} {2006})},\ \bibinfo
  {note} {ultra-accurate mass spectrometry and related topics dedicated to
  H.-J. Kluge on the occasion of his 65th birthday anniversary}\BibitemShut
  {NoStop}%
\bibitem [{\citenamefont {Savard}\ \emph {et~al.}(2008)\citenamefont {Savard},
  \citenamefont {Baker}, \citenamefont {Davids}, \citenamefont {Levand},
  \citenamefont {Moore}, \citenamefont {Pardo}, \citenamefont {Vondrasek},
  \citenamefont {Zabransky},\ and\ \citenamefont {Zinkann}}]{CARIBU}%
  \BibitemOpen
  \bibfield  {author} {\bibinfo {author} {\bibfnamefont {G.}~\bibnamefont
  {Savard}}, \bibinfo {author} {\bibfnamefont {S.}~\bibnamefont {Baker}},
  \bibinfo {author} {\bibfnamefont {C.}~\bibnamefont {Davids}}, \bibinfo
  {author} {\bibfnamefont {A.}~\bibnamefont {Levand}}, \bibinfo {author}
  {\bibfnamefont {E.}~\bibnamefont {Moore}}, \bibinfo {author} {\bibfnamefont
  {R.}~\bibnamefont {Pardo}}, \bibinfo {author} {\bibfnamefont
  {R.}~\bibnamefont {Vondrasek}}, \bibinfo {author} {\bibfnamefont
  {B.}~\bibnamefont {Zabransky}}, \ and\ \bibinfo {author} {\bibfnamefont
  {G.}~\bibnamefont {Zinkann}},\ }\href {\doibase
  https://doi.org/10.1016/j.nimb.2008.05.091} {\bibfield  {journal} {\bibinfo
  {journal} {Nuclear Instruments and Methods in Physics Research Section B:
  Beam Interactions with Materials and Atoms}\ }\textbf {\bibinfo {volume}
  {266}},\ \bibinfo {pages} {4086} (\bibinfo {year} {2008})},\ \bibinfo {note}
  {proceedings of the XVth International Conference on Electromagnetic Isotope
  Separators and Techniques Related to their Applications}\BibitemShut
  {NoStop}%
\bibitem [{\citenamefont {Davids}\ and\ \citenamefont
  {Peterson}(2008)}]{Davids08}%
  \BibitemOpen
  \bibfield  {author} {\bibinfo {author} {\bibfnamefont {C.~N.}\ \bibnamefont
  {Davids}}\ and\ \bibinfo {author} {\bibfnamefont {D.}~\bibnamefont
  {Peterson}},\ }\href {\doibase https://doi.org/10.1016/j.nimb.2008.05.148}
  {\bibfield  {journal} {\bibinfo  {journal} {Nuclear Instruments and Methods
  in Physics Research Section B: Beam Interactions with Materials and Atoms}\
  }\textbf {\bibinfo {volume} {266}},\ \bibinfo {pages} {4449} (\bibinfo {year}
  {2008})},\ \bibinfo {note} {proceedings of the XVth International Conference
  on Electromagnetic Isotope Separators and Techniques Related to their
  Applications}\BibitemShut {NoStop}%
\bibitem [{\citenamefont {Hirsh}\ \emph {et~al.}(2016)\citenamefont {Hirsh},
  \citenamefont {Paul}, \citenamefont {Burkey}, \citenamefont {Aprahamian},
  \citenamefont {Buchinger}, \citenamefont {Caldwell}, \citenamefont {Clark},
  \citenamefont {Levand}, \citenamefont {Ying}, \citenamefont {Marley},
  \citenamefont {Morgan}, \citenamefont {Nystrom}, \citenamefont {Orford},
  \citenamefont {Galvan}, \citenamefont {Rohrer}, \citenamefont {Savard},
  \citenamefont {Sharma},\ and\ \citenamefont {Siegl}}]{CARIBU-MRTOF}%
  \BibitemOpen
  \bibfield  {author} {\bibinfo {author} {\bibfnamefont {T.~Y.}\ \bibnamefont
  {Hirsh}}, \bibinfo {author} {\bibfnamefont {N.}~\bibnamefont {Paul}},
  \bibinfo {author} {\bibfnamefont {M.}~\bibnamefont {Burkey}}, \bibinfo
  {author} {\bibfnamefont {A.}~\bibnamefont {Aprahamian}}, \bibinfo {author}
  {\bibfnamefont {F.}~\bibnamefont {Buchinger}}, \bibinfo {author}
  {\bibfnamefont {S.}~\bibnamefont {Caldwell}}, \bibinfo {author}
  {\bibfnamefont {J.~A.}\ \bibnamefont {Clark}}, \bibinfo {author}
  {\bibfnamefont {A.~F.}\ \bibnamefont {Levand}}, \bibinfo {author}
  {\bibfnamefont {L.~L.}\ \bibnamefont {Ying}}, \bibinfo {author}
  {\bibfnamefont {S.~T.}\ \bibnamefont {Marley}}, \bibinfo {author}
  {\bibfnamefont {G.~E.}\ \bibnamefont {Morgan}}, \bibinfo {author}
  {\bibfnamefont {A.}~\bibnamefont {Nystrom}}, \bibinfo {author} {\bibfnamefont
  {R.}~\bibnamefont {Orford}}, \bibinfo {author} {\bibfnamefont {A.~P.}\
  \bibnamefont {Galvan}}, \bibinfo {author} {\bibfnamefont {J.}~\bibnamefont
  {Rohrer}}, \bibinfo {author} {\bibfnamefont {G.}~\bibnamefont {Savard}},
  \bibinfo {author} {\bibfnamefont {K.~S.}\ \bibnamefont {Sharma}}, \ and\
  \bibinfo {author} {\bibfnamefont {K.}~\bibnamefont {Siegl}},\ }\href
  {\doibase https://doi.org/10.1016/j.nimb.2015.12.037} {\bibfield  {journal}
  {\bibinfo  {journal} {Nuclear Instruments and Methods in Physics Research
  Section B: Beam Interactions with Materials and Atoms}\ }\textbf {\bibinfo
  {volume} {376}},\ \bibinfo {pages} {229} (\bibinfo {year} {2016})},\ \bibinfo
  {note} {proceedings of the XVIIth International Conference on Electromagnetic
  Isotope Separators and Related Topics (EMIS2015), Grand Rapids, MI, U.S.A.,
  11-15 May 2015}\BibitemShut {NoStop}%
\bibitem [{\citenamefont {Ray}\ \emph {et~al.}(2025)\citenamefont {Ray},
  \citenamefont {Valverde}, \citenamefont {Brodeur}, \citenamefont {Buchinger},
  \citenamefont {Clark}, \citenamefont {Liu}, \citenamefont {Morgan},
  \citenamefont {Orford}, \citenamefont {Porter}, \citenamefont {Savard},
  \citenamefont {Sharma},\ and\ \citenamefont {Yan}}]{Ray2025}%
  \BibitemOpen
  \bibfield  {author} {\bibinfo {author} {\bibfnamefont {D.}~\bibnamefont
  {Ray}}, \bibinfo {author} {\bibfnamefont {A.}~\bibnamefont {Valverde}},
  \bibinfo {author} {\bibfnamefont {M.}~\bibnamefont {Brodeur}}, \bibinfo
  {author} {\bibfnamefont {F.}~\bibnamefont {Buchinger}}, \bibinfo {author}
  {\bibfnamefont {J.}~\bibnamefont {Clark}}, \bibinfo {author} {\bibfnamefont
  {B.}~\bibnamefont {Liu}}, \bibinfo {author} {\bibfnamefont {G.}~\bibnamefont
  {Morgan}}, \bibinfo {author} {\bibfnamefont {R.}~\bibnamefont {Orford}},
  \bibinfo {author} {\bibfnamefont {W.}~\bibnamefont {Porter}}, \bibinfo
  {author} {\bibfnamefont {G.}~\bibnamefont {Savard}}, \bibinfo {author}
  {\bibfnamefont {K.}~\bibnamefont {Sharma}}, \ and\ \bibinfo {author}
  {\bibfnamefont {X.}~\bibnamefont {Yan}},\ }\href {\doibase
  https://doi.org/10.1016/j.nima.2025.170433} {\bibfield  {journal} {\bibinfo
  {journal} {Nuclear Instruments and Methods in Physics Research Section A:
  Accelerators, Spectrometers, Detectors and Associated Equipment}\ }\textbf
  {\bibinfo {volume} {1076}},\ \bibinfo {pages} {170433} (\bibinfo {year}
  {2025})}\BibitemShut {NoStop}%
\bibitem [{\citenamefont {König}\ \emph {et~al.}(1995)\citenamefont {König},
  \citenamefont {Bollen}, \citenamefont {Kluge}, \citenamefont {Otto},\ and\
  \citenamefont {Szerypo}}]{KONIG199595}%
  \BibitemOpen
  \bibfield  {author} {\bibinfo {author} {\bibfnamefont {M.}~\bibnamefont
  {König}}, \bibinfo {author} {\bibfnamefont {G.}~\bibnamefont {Bollen}},
  \bibinfo {author} {\bibfnamefont {H.-J.}\ \bibnamefont {Kluge}}, \bibinfo
  {author} {\bibfnamefont {T.}~\bibnamefont {Otto}}, \ and\ \bibinfo {author}
  {\bibfnamefont {J.}~\bibnamefont {Szerypo}},\ }\href {\doibase
  https://doi.org/10.1016/0168-1176(95)04146-C} {\bibfield  {journal} {\bibinfo
   {journal} {International Journal of Mass Spectrometry and Ion Processes}\
  }\textbf {\bibinfo {volume} {142}},\ \bibinfo {pages} {95} (\bibinfo {year}
  {1995})}\BibitemShut {NoStop}%
\bibitem [{\citenamefont {Bollen}\ \emph {et~al.}(1990)\citenamefont {Bollen},
  \citenamefont {Moore}, \citenamefont {Savard},\ and\ \citenamefont
  {Stolzenberg}}]{Bollen:204719}%
  \BibitemOpen
  \bibfield  {author} {\bibinfo {author} {\bibfnamefont {G.}~\bibnamefont
  {Bollen}}, \bibinfo {author} {\bibfnamefont {R.~B.}\ \bibnamefont {Moore}},
  \bibinfo {author} {\bibfnamefont {G.}~\bibnamefont {Savard}}, \ and\ \bibinfo
  {author} {\bibfnamefont {H.}~\bibnamefont {Stolzenberg}},\ }\href {\doibase
  10.1063/1.346185} {\bibfield  {journal} {\bibinfo  {journal} {J. Appl.
  Phys.}\ }\textbf {\bibinfo {volume} {68}},\ \bibinfo {pages} {4355} (\bibinfo
  {year} {1990})}\BibitemShut {NoStop}%
\bibitem [{\citenamefont {Eliseev}\ \emph {et~al.}(2013)\citenamefont
  {Eliseev}, \citenamefont {Blaum}, \citenamefont {Block}, \citenamefont
  {Droese}, \citenamefont {Goncharov}, \citenamefont {Minaya~Ramirez},
  \citenamefont {Nesterenko}, \citenamefont {Novikov},\ and\ \citenamefont
  {Schweikhard}}]{PhysRevLett.110.082501}%
  \BibitemOpen
  \bibfield  {author} {\bibinfo {author} {\bibfnamefont {S.}~\bibnamefont
  {Eliseev}}, \bibinfo {author} {\bibfnamefont {K.}~\bibnamefont {Blaum}},
  \bibinfo {author} {\bibfnamefont {M.}~\bibnamefont {Block}}, \bibinfo
  {author} {\bibfnamefont {C.}~\bibnamefont {Droese}}, \bibinfo {author}
  {\bibfnamefont {M.}~\bibnamefont {Goncharov}}, \bibinfo {author}
  {\bibfnamefont {E.}~\bibnamefont {Minaya~Ramirez}}, \bibinfo {author}
  {\bibfnamefont {D.~A.}\ \bibnamefont {Nesterenko}}, \bibinfo {author}
  {\bibfnamefont {Y.~N.}\ \bibnamefont {Novikov}}, \ and\ \bibinfo {author}
  {\bibfnamefont {L.}~\bibnamefont {Schweikhard}},\ }\href {\doibase
  10.1103/PhysRevLett.110.082501} {\bibfield  {journal} {\bibinfo  {journal}
  {Phys. Rev. Lett.}\ }\textbf {\bibinfo {volume} {110}},\ \bibinfo {pages}
  {082501} (\bibinfo {year} {2013})}\BibitemShut {NoStop}%
\bibitem [{\citenamefont {Orford}\ \emph {et~al.}(2020)\citenamefont {Orford},
  \citenamefont {Clark}, \citenamefont {Savard}, \citenamefont {Aprahamian},
  \citenamefont {Buchinger}, \citenamefont {Burkey}, \citenamefont {Gorelov},
  \citenamefont {Klimes}, \citenamefont {Morgan}, \citenamefont {Nystrom},
  \citenamefont {Porter}, \citenamefont {Ray},\ and\ \citenamefont
  {Sharma}}]{Orford2020}%
  \BibitemOpen
  \bibfield  {author} {\bibinfo {author} {\bibfnamefont {R.}~\bibnamefont
  {Orford}}, \bibinfo {author} {\bibfnamefont {J.}~\bibnamefont {Clark}},
  \bibinfo {author} {\bibfnamefont {G.}~\bibnamefont {Savard}}, \bibinfo
  {author} {\bibfnamefont {A.}~\bibnamefont {Aprahamian}}, \bibinfo {author}
  {\bibfnamefont {F.}~\bibnamefont {Buchinger}}, \bibinfo {author}
  {\bibfnamefont {M.}~\bibnamefont {Burkey}}, \bibinfo {author} {\bibfnamefont
  {D.}~\bibnamefont {Gorelov}}, \bibinfo {author} {\bibfnamefont
  {J.}~\bibnamefont {Klimes}}, \bibinfo {author} {\bibfnamefont
  {G.}~\bibnamefont {Morgan}}, \bibinfo {author} {\bibfnamefont
  {A.}~\bibnamefont {Nystrom}}, \bibinfo {author} {\bibfnamefont
  {W.}~\bibnamefont {Porter}}, \bibinfo {author} {\bibfnamefont
  {D.}~\bibnamefont {Ray}}, \ and\ \bibinfo {author} {\bibfnamefont
  {K.}~\bibnamefont {Sharma}},\ }\href {\doibase
  https://doi.org/10.1016/j.nimb.2019.04.016} {\bibfield  {journal} {\bibinfo
  {journal} {Nuclear Instruments and Methods in Physics Research Section B:
  Beam Interactions with Materials and Atoms}\ }\textbf {\bibinfo {volume}
  {463}},\ \bibinfo {pages} {491} (\bibinfo {year} {2020})}\BibitemShut
  {NoStop}%
\bibitem [{\citenamefont {Wang}\ \emph {et~al.}(2021)\citenamefont {Wang},
  \citenamefont {Huang}, \citenamefont {Kondev}, \citenamefont {Audi},\ and\
  \citenamefont {Naimi}}]{AME2020}%
  \BibitemOpen
  \bibfield  {author} {\bibinfo {author} {\bibfnamefont {M.}~\bibnamefont
  {Wang}}, \bibinfo {author} {\bibfnamefont {W.}~\bibnamefont {Huang}},
  \bibinfo {author} {\bibfnamefont {F.}~\bibnamefont {Kondev}}, \bibinfo
  {author} {\bibfnamefont {G.}~\bibnamefont {Audi}}, \ and\ \bibinfo {author}
  {\bibfnamefont {S.}~\bibnamefont {Naimi}},\ }\href {\doibase
  10.1088/1674-1137/abddaf} {\bibfield  {journal} {\bibinfo  {journal} {Chinese
  Physics C}\ }\textbf {\bibinfo {volume} {45}},\ \bibinfo {pages} {030003}
  (\bibinfo {year} {2021})}\BibitemShut {NoStop}%
\bibitem [{\citenamefont {Liu}\ \emph {et~al.}(2025)\citenamefont {Liu},
  \citenamefont {Brodeur}, \citenamefont {Clark}, \citenamefont {Dedes},
  \citenamefont {Dudek}, \citenamefont {Kondev}, \citenamefont {Ray},
  \citenamefont {Savard}, \citenamefont {Valverde}, \citenamefont {Baran},
  \citenamefont {Burdette}, \citenamefont {Houff}, \citenamefont {Orford},
  \citenamefont {Porter}, \citenamefont {Rivero}, \citenamefont {Sharma},\ and\
  \citenamefont {Varriano}}]{Liu2025}%
  \BibitemOpen
  \bibfield  {author} {\bibinfo {author} {\bibfnamefont {B.}~\bibnamefont
  {Liu}}, \bibinfo {author} {\bibfnamefont {M.}~\bibnamefont {Brodeur}},
  \bibinfo {author} {\bibfnamefont {J.~A.}\ \bibnamefont {Clark}}, \bibinfo
  {author} {\bibfnamefont {I.}~\bibnamefont {Dedes}}, \bibinfo {author}
  {\bibfnamefont {J.}~\bibnamefont {Dudek}}, \bibinfo {author} {\bibfnamefont
  {F.~G.}\ \bibnamefont {Kondev}}, \bibinfo {author} {\bibfnamefont
  {D.}~\bibnamefont {Ray}}, \bibinfo {author} {\bibfnamefont {G.}~\bibnamefont
  {Savard}}, \bibinfo {author} {\bibfnamefont {A.~A.}\ \bibnamefont
  {Valverde}}, \bibinfo {author} {\bibfnamefont {A.}~\bibnamefont {Baran}},
  \bibinfo {author} {\bibfnamefont {D.~P.}\ \bibnamefont {Burdette}}, \bibinfo
  {author} {\bibfnamefont {A.~M.}\ \bibnamefont {Houff}}, \bibinfo {author}
  {\bibfnamefont {R.}~\bibnamefont {Orford}}, \bibinfo {author} {\bibfnamefont
  {W.~S.}\ \bibnamefont {Porter}}, \bibinfo {author} {\bibfnamefont
  {F.}~\bibnamefont {Rivero}}, \bibinfo {author} {\bibfnamefont {K.~S.}\
  \bibnamefont {Sharma}}, \ and\ \bibinfo {author} {\bibfnamefont
  {L.}~\bibnamefont {Varriano}},\ }\href {\doibase 10.1103/PhysRevC.111.034308}
  {\bibfield  {journal} {\bibinfo  {journal} {Phys. Rev. C}\ }\textbf {\bibinfo
  {volume} {111}},\ \bibinfo {pages} {034308} (\bibinfo {year}
  {2025})}\BibitemShut {NoStop}%
\bibitem [{\citenamefont {Liu}(2025)}]{Liu2025Phd}%
  \BibitemOpen
  \bibfield  {author} {\bibinfo {author} {\bibfnamefont {B.}~\bibnamefont
  {Liu}},\ }\emph {\bibinfo {title} {Towards High-Precision Mass Measurements
  at the N=126 Factory}},\ \href {https://doi.org/10.7274/28766600} {\bibinfo
  {type} {Phd thesis}},\ \bibinfo  {school} {University of Notre Dame}
  (\bibinfo {year} {2025})\BibitemShut {NoStop}%
\bibitem [{\citenamefont {Kimura}\ \emph {et~al.}(2025)\citenamefont {Kimura},
  \citenamefont {Wada}, \citenamefont {Haba}, \citenamefont {Hirayama},
  \citenamefont {Ishiyama}, \citenamefont {Ito}, \citenamefont {Niwase},
  \citenamefont {Rosenbusch}, \citenamefont {Schury}, \citenamefont {Ueno},
  \citenamefont {Watanabe},\ and\ \citenamefont {Yamanouchi}}]{Kimura2025}%
  \BibitemOpen
  \bibfield  {author} {\bibinfo {author} {\bibfnamefont {S.}~\bibnamefont
  {Kimura}}, \bibinfo {author} {\bibfnamefont {M.}~\bibnamefont {Wada}},
  \bibinfo {author} {\bibfnamefont {H.}~\bibnamefont {Haba}}, \bibinfo {author}
  {\bibfnamefont {Y.}~\bibnamefont {Hirayama}}, \bibinfo {author}
  {\bibfnamefont {H.}~\bibnamefont {Ishiyama}}, \bibinfo {author}
  {\bibfnamefont {Y.}~\bibnamefont {Ito}}, \bibinfo {author} {\bibfnamefont
  {T.}~\bibnamefont {Niwase}}, \bibinfo {author} {\bibfnamefont
  {M.}~\bibnamefont {Rosenbusch}}, \bibinfo {author} {\bibfnamefont
  {P.}~\bibnamefont {Schury}}, \bibinfo {author} {\bibfnamefont
  {H.}~\bibnamefont {Ueno}}, \bibinfo {author} {\bibfnamefont {Y.~X.}\
  \bibnamefont {Watanabe}}, \ and\ \bibinfo {author} {\bibfnamefont
  {Y.}~\bibnamefont {Yamanouchi}},\ }\href {https://arxiv.org/abs/2511.11033}
  {\enquote {\bibinfo {title} {New ground state in ${}^{149}$la: Lanthanum
  isotopes lose their distinctive feature in two-neutron separation
  energies},}\ } (\bibinfo {year} {2025}),\ \Eprint
  {http://arxiv.org/abs/2511.11033} {arXiv:2511.11033 [nucl-ex]} \BibitemShut
  {NoStop}%
\bibitem [{\citenamefont {Shibata}\ \emph {et~al.}(2003)\citenamefont
  {Shibata}, \citenamefont {Shindou}, \citenamefont {Kawade}, \citenamefont
  {Kojima}, \citenamefont {Taniguchi}, \citenamefont {Kawase},\ and\
  \citenamefont {Ichikawa}}]{02ShB}%
  \BibitemOpen
  \bibfield  {author} {\bibinfo {author} {\bibfnamefont {M.}~\bibnamefont
  {Shibata}}, \bibinfo {author} {\bibfnamefont {T.}~\bibnamefont {Shindou}},
  \bibinfo {author} {\bibfnamefont {K.}~\bibnamefont {Kawade}}, \bibinfo
  {author} {\bibfnamefont {Y.}~\bibnamefont {Kojima}}, \bibinfo {author}
  {\bibfnamefont {A.}~\bibnamefont {Taniguchi}}, \bibinfo {author}
  {\bibfnamefont {Y.}~\bibnamefont {Kawase}}, \ and\ \bibinfo {author}
  {\bibfnamefont {S.}~\bibnamefont {Ichikawa}},\ }in\ \href@noop {} {\emph
  {\bibinfo {booktitle} {Exotic Nuclei and Atomic Masses}}},\ \bibinfo {editor}
  {edited by\ \bibinfo {editor} {\bibfnamefont {J.}~\bibnamefont
  {{\"A}yst{\"o}}}, \bibinfo {editor} {\bibfnamefont {P.}~\bibnamefont
  {Dendooven}}, \bibinfo {editor} {\bibfnamefont {A.}~\bibnamefont {Jokinen}},
  \ and\ \bibinfo {editor} {\bibfnamefont {M.}~\bibnamefont {Leino}}}\
  (\bibinfo  {publisher} {Springer Berlin Heidelberg},\ \bibinfo {address}
  {Berlin, Heidelberg},\ \bibinfo {year} {2003})\ pp.\ \bibinfo {pages}
  {479--479}\BibitemShut {NoStop}%
\bibitem [{\citenamefont {Graefenstedt}\ \emph {et~al.}(1987)\citenamefont
  {Graefenstedt}, \citenamefont {Keyser}, \citenamefont {M{\"u}nnich},\ and\
  \citenamefont {Schreiber}}]{87GrA}%
  \BibitemOpen
  \bibfield  {author} {\bibinfo {author} {\bibfnamefont {M.}~\bibnamefont
  {Graefenstedt}}, \bibinfo {author} {\bibfnamefont {U.}~\bibnamefont
  {Keyser}}, \bibinfo {author} {\bibfnamefont {F.}~\bibnamefont {M{\"u}nnich}},
  \ and\ \bibinfo {author} {\bibfnamefont {F.}~\bibnamefont {Schreiber}},\ }in\
  \href@noop {} {\emph {\bibinfo {booktitle} {AIP Conference Proceedings}}},\
  Vol.\ \bibinfo {volume} {164}\ (\bibinfo {organization} {American Institute
  of Physics},\ \bibinfo {year} {1987})\ pp.\ \bibinfo {pages}
  {30--40}\BibitemShut {NoStop}%
\bibitem [{\citenamefont {Ikuta}\ \emph {et~al.}(1995)\citenamefont {Ikuta},
  \citenamefont {Taniguchi}, \citenamefont {Yamamoto}, \citenamefont {Kawade},\
  and\ \citenamefont {Kawase}}]{95Ik03}%
  \BibitemOpen
  \bibfield  {author} {\bibinfo {author} {\bibfnamefont {T.}~\bibnamefont
  {Ikuta}}, \bibinfo {author} {\bibfnamefont {A.}~\bibnamefont {Taniguchi}},
  \bibinfo {author} {\bibfnamefont {H.}~\bibnamefont {Yamamoto}}, \bibinfo
  {author} {\bibfnamefont {K.}~\bibnamefont {Kawade}}, \ and\ \bibinfo {author}
  {\bibfnamefont {Y.}~\bibnamefont {Kawase}},\ }\href
  {https://api.semanticscholar.org/CorpusID:120614456} {\bibfield  {journal}
  {\bibinfo  {journal} {Journal of the Physical Society of Japan}\ }\textbf
  {\bibinfo {volume} {64}},\ \bibinfo {pages} {3244} (\bibinfo {year}
  {1995})}\BibitemShut {NoStop}%
\bibitem [{\citenamefont {{Van Klinken}}\ and\ \citenamefont
  {Taff}(1967)}]{67Va14}%
  \BibitemOpen
  \bibfield  {author} {\bibinfo {author} {\bibfnamefont {J.}~\bibnamefont {{Van
  Klinken}}}\ and\ \bibinfo {author} {\bibfnamefont {L.}~\bibnamefont {Taff}},\
  }\href {\doibase https://doi.org/10.1016/0375-9474(67)90945-1} {\bibfield
  {journal} {\bibinfo  {journal} {Nuclear Physics A}\ }\textbf {\bibinfo
  {volume} {99}},\ \bibinfo {pages} {473} (\bibinfo {year} {1967})}\BibitemShut
  {NoStop}%
\bibitem [{\citenamefont {Orford}(2019)}]{Orford2019Phd}%
  \BibitemOpen
  \bibfield  {author} {\bibinfo {author} {\bibfnamefont {R.}~\bibnamefont
  {Orford}},\ }\emph {\bibinfo {title} {A phase-imaging technique for precision
  mass measurements of neutron-rich nuclei with the Canadian Penning Trap mass
  spectrometer}},\ \href
  {https://escholarship.mcgill.ca/concern/theses/6h440v95x} {\bibinfo {type}
  {Phd thesis}},\ \bibinfo  {school} {McGill University} (\bibinfo {year}
  {2019})\BibitemShut {NoStop}%
\bibitem [{\citenamefont {Orford}\ \emph {et~al.}(2022)\citenamefont {Orford},
  \citenamefont {Vassh}, \citenamefont {Clark}, \citenamefont {McLaughlin},
  \citenamefont {Mumpower}, \citenamefont {Ray}, \citenamefont {Savard},
  \citenamefont {Surman}, \citenamefont {Buchinger}, \citenamefont {Burdette},
  \citenamefont {Burkey}, \citenamefont {Gorelov}, \citenamefont {Klimes},
  \citenamefont {Porter}, \citenamefont {Sharma}, \citenamefont {Valverde},
  \citenamefont {Varriano},\ and\ \citenamefont {Yan}}]{Orford2022}%
  \BibitemOpen
  \bibfield  {author} {\bibinfo {author} {\bibfnamefont {R.}~\bibnamefont
  {Orford}}, \bibinfo {author} {\bibfnamefont {N.}~\bibnamefont {Vassh}},
  \bibinfo {author} {\bibfnamefont {J.~A.}\ \bibnamefont {Clark}}, \bibinfo
  {author} {\bibfnamefont {G.~C.}\ \bibnamefont {McLaughlin}}, \bibinfo
  {author} {\bibfnamefont {M.~R.}\ \bibnamefont {Mumpower}}, \bibinfo {author}
  {\bibfnamefont {D.}~\bibnamefont {Ray}}, \bibinfo {author} {\bibfnamefont
  {G.}~\bibnamefont {Savard}}, \bibinfo {author} {\bibfnamefont
  {R.}~\bibnamefont {Surman}}, \bibinfo {author} {\bibfnamefont
  {F.}~\bibnamefont {Buchinger}}, \bibinfo {author} {\bibfnamefont {D.~P.}\
  \bibnamefont {Burdette}}, \bibinfo {author} {\bibfnamefont {M.~T.}\
  \bibnamefont {Burkey}}, \bibinfo {author} {\bibfnamefont {D.~A.}\
  \bibnamefont {Gorelov}}, \bibinfo {author} {\bibfnamefont {J.~W.}\
  \bibnamefont {Klimes}}, \bibinfo {author} {\bibfnamefont {W.~S.}\
  \bibnamefont {Porter}}, \bibinfo {author} {\bibfnamefont {K.~S.}\
  \bibnamefont {Sharma}}, \bibinfo {author} {\bibfnamefont {A.~A.}\
  \bibnamefont {Valverde}}, \bibinfo {author} {\bibfnamefont {L.}~\bibnamefont
  {Varriano}}, \ and\ \bibinfo {author} {\bibfnamefont {X.~L.}\ \bibnamefont
  {Yan}},\ }\href {\doibase 10.1103/PhysRevC.105.L052802} {\bibfield  {journal}
  {\bibinfo  {journal} {Phys. Rev. C}\ }\textbf {\bibinfo {volume} {105}},\
  \bibinfo {pages} {L052802} (\bibinfo {year} {2022})}\BibitemShut {NoStop}%
\bibitem [{\citenamefont {Ray}\ \emph {et~al.}(2024)\citenamefont {Ray},
  \citenamefont {Vassh}, \citenamefont {Liu}, \citenamefont {Valverde},
  \citenamefont {Brodeur}, \citenamefont {Clark}, \citenamefont {McLaughlin},
  \citenamefont {Mumpower}, \citenamefont {Orford}, \citenamefont {Porter},
  \citenamefont {Savard}, \citenamefont {Sharma}, \citenamefont {Surman},
  \citenamefont {Buchinger}, \citenamefont {Burdette}, \citenamefont
  {Callahan}, \citenamefont {Gallant}, \citenamefont {Hoff}, \citenamefont
  {Kolos}, \citenamefont {Kondev}, \citenamefont {Morgan}, \citenamefont
  {Rivero}, \citenamefont {Santiago-Gonzalez}, \citenamefont {Scielzo},
  \citenamefont {Varriano}, \citenamefont {Weber}, \citenamefont {Wilson},\
  and\ \citenamefont {Yan}}]{Ray2024}%
  \BibitemOpen
  \bibfield  {author} {\bibinfo {author} {\bibfnamefont {D.}~\bibnamefont
  {Ray}}, \bibinfo {author} {\bibfnamefont {N.}~\bibnamefont {Vassh}}, \bibinfo
  {author} {\bibfnamefont {B.}~\bibnamefont {Liu}}, \bibinfo {author}
  {\bibfnamefont {A.~A.}\ \bibnamefont {Valverde}}, \bibinfo {author}
  {\bibfnamefont {M.}~\bibnamefont {Brodeur}}, \bibinfo {author} {\bibfnamefont
  {J.~A.}\ \bibnamefont {Clark}}, \bibinfo {author} {\bibfnamefont {G.~C.}\
  \bibnamefont {McLaughlin}}, \bibinfo {author} {\bibfnamefont {M.~R.}\
  \bibnamefont {Mumpower}}, \bibinfo {author} {\bibfnamefont {R.}~\bibnamefont
  {Orford}}, \bibinfo {author} {\bibfnamefont {W.~S.}\ \bibnamefont {Porter}},
  \bibinfo {author} {\bibfnamefont {G.}~\bibnamefont {Savard}}, \bibinfo
  {author} {\bibfnamefont {K.~S.}\ \bibnamefont {Sharma}}, \bibinfo {author}
  {\bibfnamefont {R.}~\bibnamefont {Surman}}, \bibinfo {author} {\bibfnamefont
  {F.}~\bibnamefont {Buchinger}}, \bibinfo {author} {\bibfnamefont {D.~P.}\
  \bibnamefont {Burdette}}, \bibinfo {author} {\bibfnamefont {N.}~\bibnamefont
  {Callahan}}, \bibinfo {author} {\bibfnamefont {A.~T.}\ \bibnamefont
  {Gallant}}, \bibinfo {author} {\bibfnamefont {D.~E.~M.}\ \bibnamefont
  {Hoff}}, \bibinfo {author} {\bibfnamefont {K.}~\bibnamefont {Kolos}},
  \bibinfo {author} {\bibfnamefont {F.~G.}\ \bibnamefont {Kondev}}, \bibinfo
  {author} {\bibfnamefont {G.~E.}\ \bibnamefont {Morgan}}, \bibinfo {author}
  {\bibfnamefont {F.}~\bibnamefont {Rivero}}, \bibinfo {author} {\bibfnamefont
  {D.}~\bibnamefont {Santiago-Gonzalez}}, \bibinfo {author} {\bibfnamefont
  {N.~D.}\ \bibnamefont {Scielzo}}, \bibinfo {author} {\bibfnamefont
  {L.}~\bibnamefont {Varriano}}, \bibinfo {author} {\bibfnamefont {C.~M.}\
  \bibnamefont {Weber}}, \bibinfo {author} {\bibfnamefont {G.~E.}\ \bibnamefont
  {Wilson}}, \ and\ \bibinfo {author} {\bibfnamefont {X.~L.}\ \bibnamefont
  {Yan}},\ }\href {https://arxiv.org/abs/2411.06310} {\enquote {\bibinfo
  {title} {Mass measurements of neutron-rich nuclides using the canadian
  penning trap to inform predictions in the $r$-process rare-earth peak
  region},}\ } (\bibinfo {year} {2024}),\ \Eprint
  {http://arxiv.org/abs/2411.06310} {arXiv:2411.06310 [nucl-ex]} \BibitemShut
  {NoStop}%
\end{thebibliography}
\end{document}